\documentclass[aps,amssymb,amsmath,preprint,showpacs]{revtex4}
\usepackage{graphicx}
\newcommand{\ket}[1]{\left|{#1}\right\rangle}
\newcommand{\modu}[1]{\left|{#1}\right|}
\newcommand{\expect}[3]{\left\langle{#1}\right|{#2}\left|{#3}\right\rangle}
\newcommand{\aver}[1]{\left\langle{#1}\right\rangle}
\newcommand{\added}[2]{\left|{#1},{#2}\right\rangle}

\newcommand{\beq}{\begin{equation}}
\newcommand{\eeq}{\end{equation}}
\begin{document}
%\preprint{IITM/PH/TH/2004/3}
%\preprint{}
\title{Wave packet dynamics of the matter wave field of 
a Bose-Einstein condensate}
\author{C. Sudheesh, S. Lakshmibala, and V. Balakrishnan}
\email{sudheesh,slbala,vbalki@physics.iitm.ac.in}
\affiliation{
Department of Physics, Indian Institute of Technology Madras,
Chennai 600 036, India}
\date{23 June 2004}
\begin{abstract}
We show in the framework of a tractable model
that revivals and fractional 
revivals of wave packets afford clear signatures of the 
extent of departure from
coherence and from Poisson statistics 
of the matter wave field in a Bose-Einstein condensate, 
or of a suitably chosen initial state of the radiation field 
propagating in a Kerr-like medium. 

\end{abstract}
\maketitle

The evolution of a quantum wave packet subject to a 
confining potential is of current interest in several 
experimentally realizable situations. Foremost among these 
is the dynamics 
of the matter wave field 
of a Bose-Einstein condensate (BEC) confined by a 
three-dimensional optical lattice\cite{grei}. 
The condensate is, in general, in a coherent 
superposition of different
atom-number states\cite{wri1,ima,cast} with a repulsive 
interaction between the atoms.
This state evolves in time in the confining potential. 
If the number of
atoms and the number of lattice sites are both sufficiently large, 
the atom number
distribution in each well obeys Poisson 
statistics to a good approximation.
With increasing well depth and decreasing tunneling energy, the 
wells can be taken to be sufficiently isolated from each other. The
atom number distribution in each well then departs from the
Poisson, and 
significant non-classical effects manifest themselves\cite{grei}. 
These include  
squeezing and
dissipation, sub-Poisson 
statistics, as well as revivals and fractional revivals at 
particular instants of time. 
A realistic and tractable model for the dynamics of
the atoms in each well is provided by the Hamiltonian 
\beq
H = \hbar\chi 
a^{\dagger 2} a^2= \hbar\chi N(N-1),
\label{hamil}
\eeq
where $a\,, a^\dagger$ and $N$ are annihilation,  
creation and number operators of atoms, and 
$\chi$ characterizes the energy needed to 
overcome the inter-atomic repulsion in adding an atom to the 
population of the potential well.

As is well known, the formal equivalence between bosonic atoms and 
photons has been exploited in recent years to bring out the deep
analogies between quantum optics and atom optics involving
BECs\cite{molm}, enabling a fruitful two-pronged approach 
to the problems of
quantum computing. Wave packets propagating in a nonlinear optical
medium display precisely the 
variety of non-classical 
features\cite{wall} mentioned above that BECs display. In particular, 
revivals and fractional revivals, which are
now recognized to be generic features of wave packet evolution in
nonlinear quantum dynamics,  constitute
a striking aspect of experimental observations\cite{grei} of BECs.
The revival phenomenon\cite{robi,park} 
 has been studied in detail in diverse
situations\cite{kita,yurk,tara,sesh1,sesh2}, including 
that of the dynamics of a single-mode field 
propagating in a Kerr-like 
medium. 
In this case the initial wave packet 
is a coherent state of 
the radiation field, and it is precisely the Hamiltonian 
in Eq. ({\ref{hamil}) (where the operators now refer to photons)
that governs the dynamics.

The identification of clear 
signatures of revivals and fractional revivals 
helps distinguish between wave packets 
that obey Poisson statistics and those that obey 
sub-Poissonian  statistics, and also provides valuable 
information on the degree of coherence enjoyed by the system. 
This would also be of practical importance  
in quantum computation using wave packets, where 
logic gate 
operations are envisaged to be implemented 
at the precise instants of fractional 
revivals\cite{shap}. 

Specifying the state of a BEC in an actual experiment is not
simple, and several models are extant. However, plausibility arguments
may be given to support a pure state description\cite{wri1} of the
BEC, according to which the state at any time $t$ 
has the general form
\begin{equation}
\ket{\psi(t)}=\sum_n \frac{b_{n}(t)}{\sqrt{n!}}
\ket{n},
\label{condensatewavefn}
\end{equation}
where the expansion coefficients $b_{n}(t)$ are model-dependent.  
It is of great interest to study the departure from 
coherence of an initial state under time evolution governed by a
hermitian but nonlinear Hamiltonian as in Eq.~(\ref{hamil}). 
What is
required for this purpose is a model initial state that has three basic
features: a precisely quantifiable, preferably tunable, 
degree of departure
from perfect coherence, sub-Poissonian statistics (a standard
deviation that is
less than the mean), and phase-squeezing. All these 
properties are possessed by the normalized state
\begin{equation}
\added{\alpha}{m}
=\frac{(a^\dagger)^m\ket{\alpha}}{\sqrt{\expect{\alpha}{a^m 
a^{\dagger m}}{\alpha}}}
=\frac{(a^\dagger)^m\ket{\alpha}}{\sqrt{
m!\,L_{m}(-\nu)}}
\label{photonadded} 
\end{equation}
where $m$ is a positive integer, $\ket{\alpha}$ is the standard
oscillator coherent 
state defined by $a\ket{\alpha}=\alpha\ket{\alpha},\,\,\alpha\in 
\mathbb{C},\,\nu \equiv |\alpha|^{2},$ and $L_{m}$ is the Laguerre
polynomial of order $m$. The departure of $\added{\alpha}{m}$ 
from perfect coherence arises 
due to the addition of $m$ atoms to $\ket{\alpha}$, a feature that 
becomes more 
pronounced  with increasing $m$. In the context of quantum optics, in
which this state has first been studied\cite{agar2},  
$\added{\alpha}{m}$ is called an $m$-photon-added coherent state, 
and is produced in laser-atom 
interactions under appropriate conditions. Its non-classical features
include both phase squeezing and
sub-Poissonian statistics. 
The latter property implies that the standard deviation in the atom
number $N$ of the state behaves like $N^{\frac{1}{2}-\beta}$ rather
than $N^{\frac{1}{2}}$, the exponent $\beta$ being a calculable
decreasing function of $m$.

While $\added{\alpha}{m}$ is not an 
eigenstate of $a$, it may be regarded\cite{siva} as a ``nonlinear coherent 
state'',  in the sense that it is an eigenstate of the 
operator $[1- m \,(1+a^\dagger a)^{-1}]\,a$ 
with eigenvalue $\alpha$.
The state $\added{\alpha}{m}$ can also be viewed in another way. Instead of
the Fock basis $\{\ket{n}\}$, we may consider the unitarily 
transformed basis $\{\added{n}{\alpha}\}$
formed by the generalized coherent states 
$\added{n}{\alpha}=\exp\,(\alpha \,a^\dagger-\alpha ^{*}\, a)\,\ket{n}.$
(Equivalently, for a given $n$, $\added{n}{\alpha}$ is simply the state    
$(a^\dagger-\alpha^{*})^{n} \ket{\alpha}$, 
normalized to unity.) The initial state $\ket
{\psi(0)}$ of the condensate can be expanded in the basis
$\{\added{n}{\alpha}\}$ instead of the Fock basis.
Likewise, it can be shown that, for a given $m$,  
the state $\added{\alpha}{m}$ is a superposition of the form $\sum_{n=0}^{m}
c_{n} \added{n}{\alpha}$.  
Thus, in practice, $\added{\alpha}{m}$ is a very appropriate candidate 
for the initial state of the condensate. 
  
In this paper we show that distinctive signatures of revivals and 
fractional revivals of the 
condensate are manifested in the mean values of certain basic  
operators pertaining to the system, and in their variances.
We examine the precise manner in which the 
departure from coherence of the initial condensate 
wave function affects its subsequent dynamics, particularly at the instants of
revivals and 
fractional revivals. The distinctions between different fractional
revivals that occur in between two successive revivals are also
brought out.

The matter wave field in a BEC is essentially given by the expectation value 
$\langle \psi (t)|a|\psi (t)\rangle$. Its real and imaginary parts 
are the expectation values of the hermitian combinations 
$(a + a^{\dagger})/2$ and $-i(a - a^{\dagger})/2$, which in turn 
correspond to the cases
$\varphi = 0$ and $\varphi = -\frac{1}{2}\pi$ of the 
field quadrature 
$\xi = (a\,e^{i\varphi} + a^{\dagger}\,e^{-i\varphi})/2$ that is 
customarily taken\cite{agar2} 
as the basic observable in the quantum optics context.
We therefore set
$x= (a+a^\dagger)/2^{1/2}$ and $p= -i(a-a^\dagger)/2^{1/2}$ (so that 
$[x\,,\,p] = i$),  
and examine the expectation values and variances of $x$ and $p$ as the
system evolves from the initial state in Eq. (\ref{photonadded}) under
the Hamiltonian $H$ of Eq. (\ref{hamil}). 
We shall see that the time dependence of these mean  
values and the corresponding variances 
mirrors, in distinct ways, the occurrence of different 
fractional revivals between 
two  successive revivals of the initial state. 
We use the convenient notation
\begin{equation}
\aver{x(t)}_{m} = 
\expect{\alpha\,,\,m}{e^{iHt/\hbar}\,x\,e^{-iHt/\hbar}}
{\alpha\,,\,m},
\label{expectm}
\end{equation}
with an analogous definition for 
$\aver{p(t)}_{m}\,.$ 
As $H$ is diagonal in the
number operator $N$, it follows that the mean atom number, 
given by\cite{agar2}
\begin{equation}
\aver{N}_{m} = \frac{(m+1)\,L_{m+1}(-\nu)}{L_{m}(-\nu)} - 1, 
\label{expectN}
\end{equation}
remains constant in time.
When $\nu = 0, \,\aver{N}_{m}$  reduces to $m,$ as required; while 
for $\nu \gg m\,,
\aver{N}_{m} = \nu + 2m + \mathcal{O}(\nu^{-1}).$ The moments of $N$,
and hence the sub-Poissonian statistics of the atom number, also
remain unaltered in time.

It is helpful to use 
as a reference, for the purposes of subsequent comparison, the results
that obtain in the case when
$m = 0$, i.e., for an initial state that is just the coherent state
$\ket{\alpha}$ (which has, of course, a Poissonian number distribution
with mean value $\nu$).  
A straightforward calculation gives 
\begin{eqnarray}
\aver{x(t)}_{0}
&=&  
\exp\,[-\nu\,(1-\cos 2\chi t)]\nonumber\\   
&\times& \big\{ x_0 \,\cos \,(\nu \sin 2\chi t)
+p_0\,\sin \,(\nu \sin 2\chi t)\big\},
\label{xt}
\end{eqnarray}
\begin{eqnarray}
\aver{p(t)}_{0}
&=&\exp\,[-\nu(1-\cos 2\chi t)]\nonumber\\
&\times& \big\{p_0 \,\cos\, (\nu \sin 2\chi t)
-x_0\, 
\sin \,(\nu \sin 2\chi t)\big\}.
\label{pt}
\end{eqnarray}
Here 
$\alpha = (x_0+ip_0)/2^{1/2}$ so that 
$\nu= \modu{\alpha}^2=\frac{1}{2}(x_0^2+p_0^2)\,.$ 
The parameters 
$x_0$ and $p_0$ signify (in the case at hand, namely, for $m = 0$) 
the locations of the centers of the initial 
Gaussian wave packets in position and momentum space, respectively.
It is evident that 
$\aver{x(t)}_0$ and $\aver{p(t)}_0$ are periodic in $t$, with 
a period 
$\pi/\chi = T_{\rm rev}\,,$ 
the revival time. (For 
the Hamiltonian 
in Eq. (\ref{hamil}), $T_{\rm rev}$  
coincides with the 
``classical orbit time'' $T_{\rm cl}$ because the
coefficients of the terms linear and quadratic in $N$ happen to be 
equal in magnitude.) In fact, at the level 
of expectation values,
this case can be cast in the form of a classical nonlinear
oscillator with ``dynamical variables''
$X_{0}\equiv\aver{x(t)}_{0}\,\exp\,[\nu(1-\cos 2\chi t)]$ and $P_{0}
\equiv\aver{p(t)}_{0} 
\,\exp\,[\nu(1-\cos 2\chi t)]:$ setting $z_{0}(t) =
\exp\,(i\nu\,\sin\,2\chi t),$ Eqs. (\ref{xt}) and (\ref{pt}) become
\begin{eqnarray}
X_{0}(t) = x_{0}\,{\rm Re}\,z_{0}(t) + p_{0}\,{\rm
  Im}\,z_{0}(t),\nonumber\\
P_{0}(t) = p_{0}\,{\rm Re}\,z_{0}(t) - x_{0}\,{\rm
  Im}\,z_{0}(t).
\label{m0xtpt}
\end{eqnarray}
If we now {\em re-parametrize} time according to $\tau =\sin\, 2\chi t,$ 
we have $ dX_0/d\tau=\nu P_0$ and $dP_0/d\tau=-\nu X_0\,.$
Thus, formally, we have essentially a nonlinear oscillator 
of frequency $\frac{1}{2}[X_{0}^{2}(0)+P_{0}^{2}(0)],\,$ 
$X_{0}(0)$ and $P_{0}(0)$ being the initial values of 
$X_{0}(\tau)$ and $P_{0}(\tau),$ 
respectively.

The time dependence of 
$\aver{x(t)}_{m}$ 
and $\aver{p(t)}_{m}$ 
for $m \neq 0$ differs in striking ways from the
foregoing, even for small values of $m.$   
The revival time $T_{\rm rev}$ 
remains equal to $\pi/\chi,$ of course, but   
in the intervals between revivals the
time evolution is considerably more involved than the
expressions in Eqs. (\ref{xt}) and (\ref{pt}) for the case $ m = 0.$ 
This implies that
even a small departure from coherence in the
initial state of the condensate and from Poissonian number statistics 
leads to a very different time 
evolution of the system and the phase squeezing it exhibits. 
We have calculated the exact expressions for $\aver{x(t)}_{m}$ 
and $\aver{p(t)}_{m}\,,$ and these are as follows.  
Their initial values are given by
\begin{equation}
\aver{x(0)}_{m} =
\frac{L_{m}^{(1)}(-\nu)}{L_{m}(-\nu)}\,x_{0}\,,\,\,
\aver{p(0)}_{m} =
\frac{L_{m}^{(1)}(-\nu)}{L_{m}(-\nu)}\,p_{0}\,,
\label{xpinitial}
\end{equation}
where $L_{m}^{(1)}(-\nu) = d\,L_{m+1}(-\nu)/d\nu$ is an 
associated Laguerre polynomial. 
Analogous to $(X_{0}\,,\,P_{0})\,,$
let us define 
\begin{eqnarray}
X_{m}(t)&=&\aver{x(t)} _{m}\,
\exp\,[\nu\,(1-\cos 2\chi t)],\nonumber\\  
P_{m}(t)&=&\aver{p(t)} _{m}\,\exp\,[\nu\,(1-\cos 2\chi t)].
\label{XPm}
\end{eqnarray} 
The solutions for these quantities may then be written in the 
compact and suggestive form   
\begin{eqnarray}
X_{m}(t)&=& x_{0}\,{\rm Re}\,z_{m}(t)+p_{0}\,{\rm Im}\,z_{m}(t),\nonumber\\
P_{m}(t)&=& p_{0}\,{\rm Re}\,z_{m}(t)-x_{0}\,{\rm Im}\,z_{m}(t),
\label{mxtpt}
\end{eqnarray}
where 
\begin{equation}
z_{m}(t)=
\frac{L_{m}^{(1)}(-\nu\,e^{2i\chi t})}{L_{m}(-\nu)}
\,\exp\,[i\,(2 m \chi t+\nu\,\sin\,2\chi t)]\,.
\label{zm}
\end{equation} 
A number of differences between these results and those for the case
$m = 0$ are noteworthy. First, $|z_{m}|$ varies with $t$, in contrast
to $|z_{0}(t)|
\equiv 1.$ (Note also that $z_{m}(0) = 
L_{m}^{(1)}(-\nu)/L_{m}(-\nu) \neq
1.$) The time dependence of $X_{m}$ and $P_{m}$ involves the sines and
cosines of the set of arguments $(2\chi \,l t + \nu\,\sin\,2\chi t)$ 
where $l = m,\ldots\,,\,2m.$ 
Thus, not only are
``higher harmonics'' present, but the arguments also involve {\em secular}
(linear) terms in $t$ added to the original $\nu\,\sin\,2\chi t.$  This
important difference precludes the possibility of 
subsuming the time dependence into that of an effective nonlinear
oscillator by means of a re-parametrization of the time, unlike 
the case $m = 0.$  

We present the rest of our results with the help of figures based on
numerical computation. In what follows, we set $\chi = 5$ for
definiteness, and also restrict ourselves to the case 
$x_0=p_0$ (there is no significant loss of generality as
a result of this symmetric choice of parameters). The presence of the
overall factor $\exp\,[-\nu\,(1-\cos 2\chi t)]$ in the expressions for 
$\aver{x(t)}_{m}$ 
and $\aver{p(t)}_{m}$ implies that, for sufficiently large values of
the parameter $\nu$, the expectation values remain essentially static
around the value zero, and burst into rapid variation
only in the neighborhood of revivals. Smaller values of $\nu$ enable us to
resolve the details of the time variation more clearly.  

Figures \ref{becfig1}(a) and (b) are, respectively,  plots of 
the expectation values 
$\aver{x(t)}_{0}$ and $\aver{x(t)}_{10}$ versus 
$t$ (in units of $T_{\rm rev}$) for parameter values  
$x_0=p_0=1$ (i.e., for $\nu=1$). The revivals at integer values of 
$t/T_{\rm rev}$ are manifest. With increasing $m$ (or a decreasing
degree of coherence in the initial state), the relatively smooth 
behavior of $\aver{x(t)}_{0}$ gives way to increasingly
rapid oscillatory behavior in the vicinity of revivals. 
The range over which the expectation value varies also
increases for larger values of $m$. 
\begin{figure} 
\includegraphics[width=3.3in,height=1.6in]{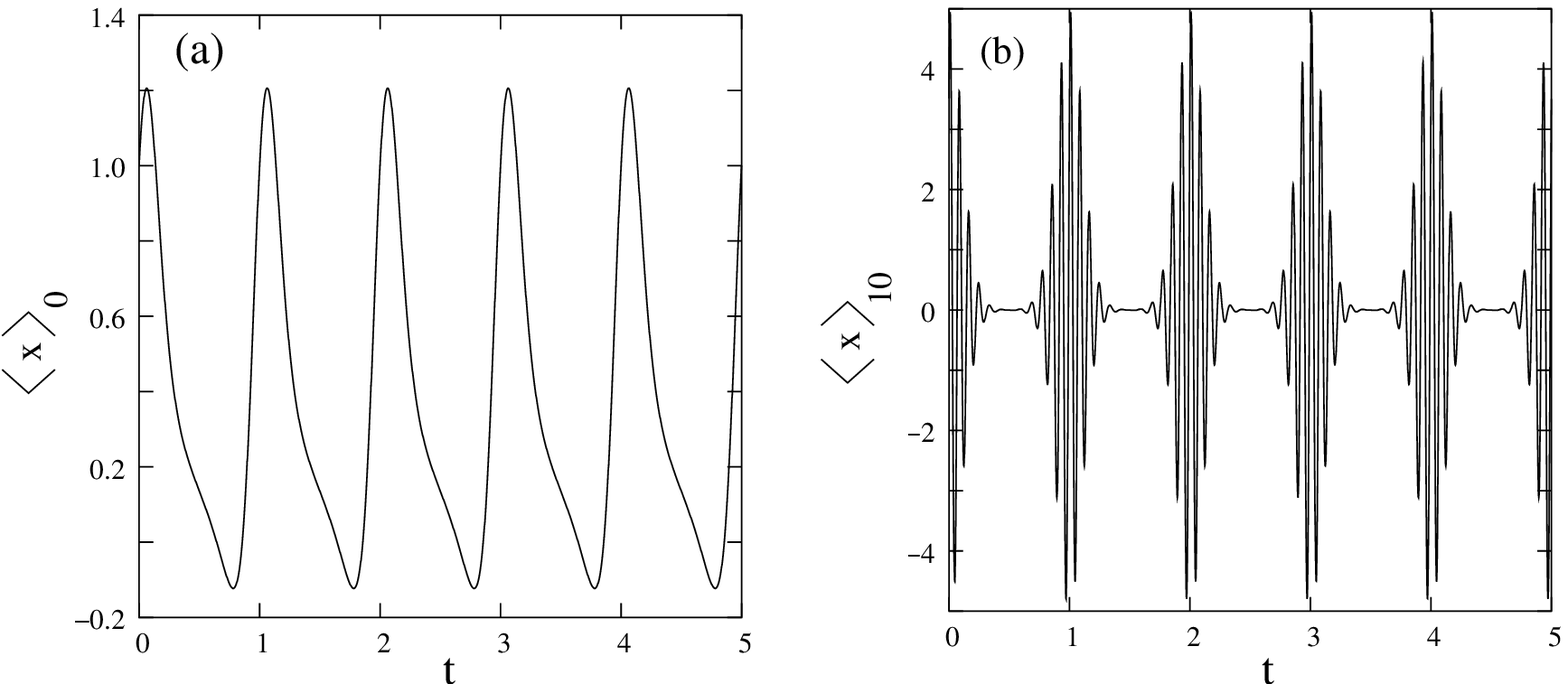} 
\caption{$\aver{x(t)}_{0}$ as a function of time}
\label{becfig1}
\end{figure}
\begin{figure} 
\includegraphics[width=3.3in,height=1.6in]{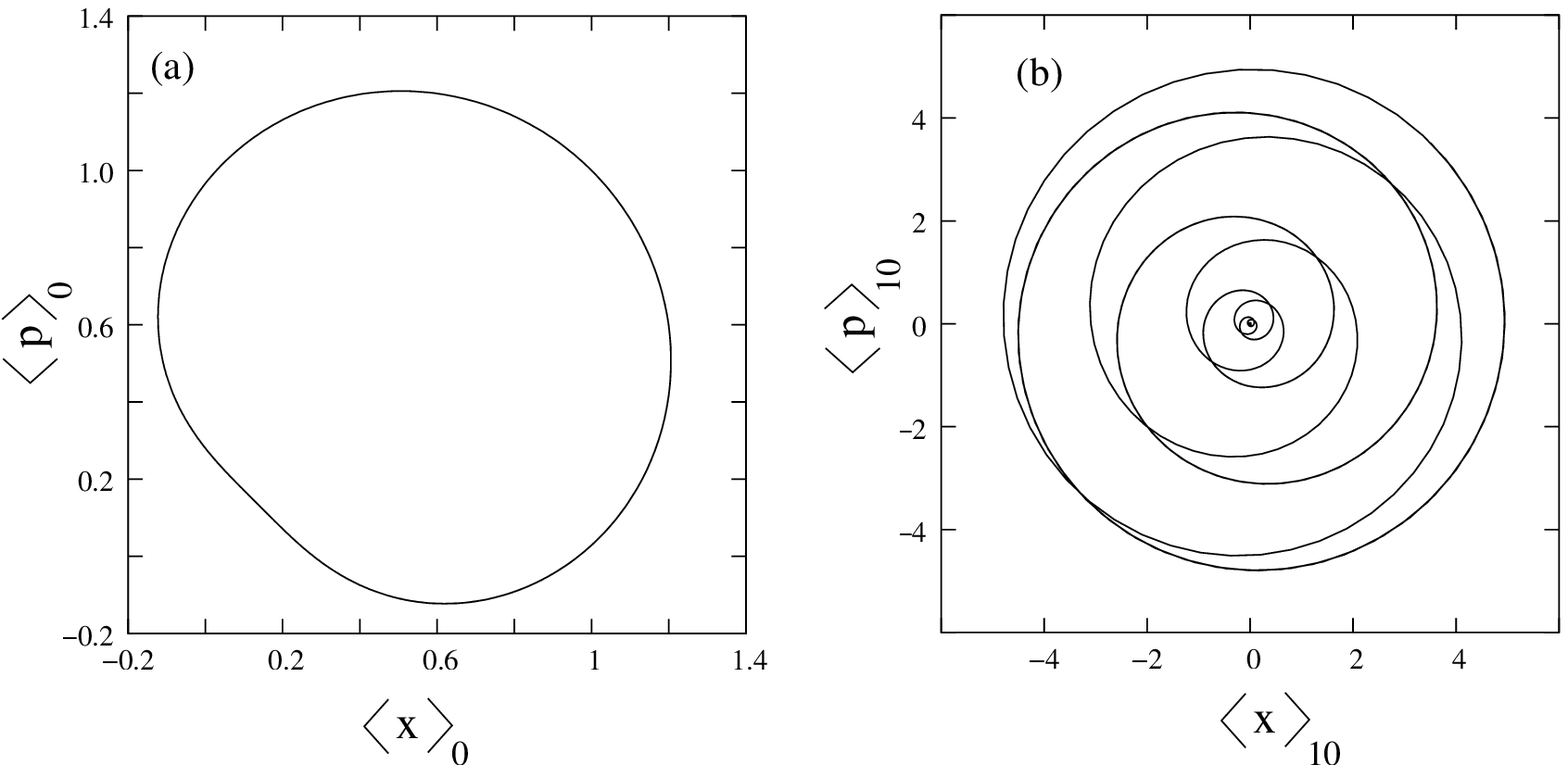} 
\caption{``Phase plot'' of $\aver{p}$ {\it vs} $\aver{x}$}
\label{becfig2}
\end{figure}
Essentially the same sort of behavior is shown by  
$\aver{p(t)}_{m}\,.$ 
However, a ``phase plot'' of 
$\aver{p(t)}_{m}$ versus $\aver{x(t)}_{m}\,$ in Figs. \ref{becfig2}(a)
and (b)  
reveals complementary aspects of
such oscillatory behavior with increasing $m$, showing how the
oscillations in the two quantities go in and out of phase with each
other. (The entire closed curve in each case is traversed in a time
period $T_{\rm rev}$.)   

The initial state also undergoes fractional revivals in the interval
between any two successive revivals. In principle, it can be shown
that, in the interval $(0\,,\,T_{\rm rev}),$ fractional revivals occur
at the instants 
$T_{\rm rev}/k$ 
where $k = 2,3,\ldots\,,$ and also, for any
given $k$, at the instants 
$j\,T_{\rm rev}/k$ 
where $j =
1,2,\ldots\,,(k-1).$ 
These fractional revivals are signaled by the appearance
of  $k$ spatially-distributed wave packets similar to the wave
packet representing the state at $t = 0.$ 
The fractional revivals at 
the instants $j\,T_{\rm rev}/k$ show up 
in the rapid pulsed variation of the 
$k^{\rm th}$ moments of $x$ and $p,$ and not in the lower
moments\cite{sudh}. However, if we use  
$\added{\alpha}{m}$ as the initial state, then, even for relatively
small values of $m$, the signatures of fractional revivals appear  
for values of $\nu$
that are not large, in contrast to what happens when the initial state
is the coherent state $\ket{\alpha}.$ 
(Recall that $\aver{N}_{m}$ 
is determined by $\nu$ according
to Eq. (\ref{expectN}).) 

For illustrative purposes we investigate the specific case of the 
fractional revival at 
$t=\frac{1}{2}T_{\rm rev}\,.$ 
This corresponds to the appearance of 
two spatially separated similar wave packets, i.e., a single 
qubit in the language 
of logic gate operations. Plots of the product 
$\Delta x\, \Delta p$ of the standard deviations of $x$ and $p$ 
versus $t/T_{\rm rev}$ over a full cycle  
are shown in 
Fig. \ref{becfig3} for initial states given, respectively, by  
the coherent state $\ket{\alpha}$ (dotted curve), the atom-added
state $\added{\alpha}{1}$ (dashed curve),
and  the multi-atom-added state $\added{\alpha}{10}$\, (bold curve), 
for the same parameter
values as above ($\chi=5,\,x_0=p_0=1,$ so that $\nu = 1.$)
\begin{figure} 
\includegraphics[width=2in,height=2in]{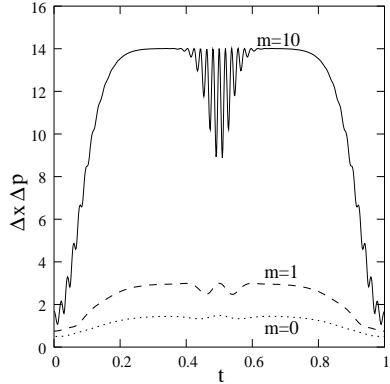} 
\caption{Variation of the uncertainty product with time}
\label{becfig3}
\end{figure}
It is seen that hardly any trace of the fractional revival is 
evident 
in the case of an initially coherent state, in marked contrast to the case of
the atom-added states, in which 
the fractional revival is signaled by oscillations 
whose frequency and amplitude increase quite rapidly with increasing
$m$. This effect gets masked for 
larger values of the parameter $\nu$,
when these oscillations are relatively insensitive to the value of
$m$.  
%This is evident in Fig. 4, which is plotted for $\nu = 100$ in
%the cases $m = 0$ (dotted line), $m = 1$ 
%(dashed line) and $m = 10$ (bold line), respectively. 

Another striking feature that provides a clear distinction 
between the revivals at
$t = n\,T_{\rm rev}$ and the fractional revivals at $t =
(n+\frac{1}{2})\,T_{\rm rev}$ is illustrated in Fig. \ref{becfig4ab}
(a), which is a
plot of $\Delta p$ versus $\Delta x\,.$ The dotted and full lines
correspond to $m = 0$ and $m = 5,$ respectively. 
At $t = 0$, these quantities
are equal, and have small values. 
\begin{figure} 
\includegraphics[width=3.3in,height=1.75in]{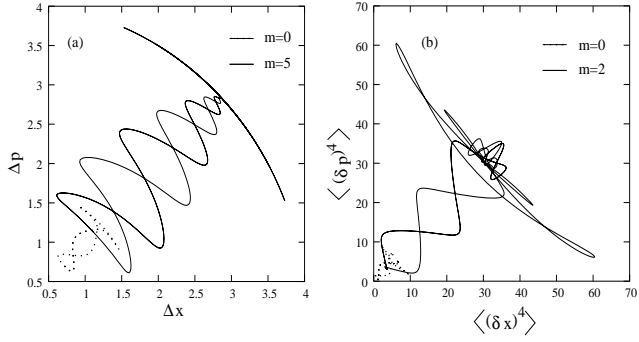} 
\caption{``Phase plots'' of higher moments of $p$ and $x$}
%\caption{$\Delta p$ {\em vs} $\Delta x$}
\label{becfig4ab}
\end{figure}
As $t$ increases, they rapidly build
up, oscillating about the radial $\Delta p = \Delta x$ line with an
initially increasing, and then decreasing, amplitude. A maximum value
of $\Delta x$ and $\Delta p$ is attained, at which these quantities 
then remain nearly static, till the onset of 
the fractional revival at $T_{\rm rev}\,.$ They then begin to 
oscillate rapidly once again, but this time
in a {\em tangential} direction, swinging back and forth along an arc 
with an amplitude that initially increases and then decreases to zero:
in other words, the individual standard deviations fluctuate 
rapidly in the
vicinity of the fractional revival (while $[(\Delta x)^2 + (\Delta
p)^2]^{1/2}$ remains esentially unchanged in magnitude), 
in marked contrast to what happens at a
revival. It is evident that all these features are very 
significantly
enhanced and magnified for non-zero values of $m$, 
relative to what happens for the
case $m = 0.$   

Similar signatures of fractional revivals for higher values of $k$ 
can be discerned by using initial states $\added{\alpha}{m}$ even with
relatively small values of $\nu$, the  
oscillations in the moments of observables becoming more pronounced
with increasing $m$. For instance, signatures of the 
fractional revivals at $t = 
\frac{1}{4}j\,T_{\rm rev}\,,\,\,j = 1,\,2,\,3$ are clearly discernible
in the behavior of the fourth moments of $x.$
%\begin{figure} 
%\includegraphics[width=2.5in,height=1.75in]{becfig5.eps} 
%\caption{$\aver{(\delta p)^4}$ {\em vs} $\aver{(\delta x)^4}$} 
%\label{becfig5}
%\end{figure}
Figure \ref{becfig4ab} (b) is a plot of 
$\aver{(\delta p)^4}$ 
versus $\aver{(\delta x)^4}$ where $\delta x = x - \aver{x}\,,\,\,
\delta p = p - \aver{p},$ for $\nu = 1.$ The dotted and full lines
correspond to the initial states $\ket{\alpha}$ and $\added{\alpha}{2}$
respectively. Once again, the magnification of the variations  
that occurs for even a
small value of $m$ is manifest.

In conclusion, we have demonstrated that
an atom-added initial state of the form  
$\added{\alpha}{m}$ shows significantly 
increased sensitivity to revivals and
fractional revivals. In this sense, one may therefore expect that  
the inevitable departure, in practice, 
of the initial state of a BEC
from perfect coherence 
can in fact be used to advantage.

We thank P. K. Panigrahi for discussions. This work was 
supported in part by the 
Department of Science and Technology, India, under Project 
No. SP/S2/K-14/2000.

\end{document}